\documentclass[10pt,a4paper]{article}

\usepackage[a4paper,top=20mm,bottom=20mm,left=21mm,right=21mm]{geometry}
\usepackage{mathptmx}

\usepackage{amsmath,amssymb,amsfonts,mathtools}
\usepackage{bm}
\usepackage{slashed}
\usepackage{graphicx}
\usepackage{booktabs}
\usepackage{tabularx}
\usepackage{caption}
\usepackage{subcaption}
\usepackage[compress]{cite}
\usepackage{xcolor}
\usepackage{fancyhdr}
\usepackage[colorlinks=true,linkcolor=blue!70!black,citecolor=blue!70!black,urlcolor=blue!70!black]{hyperref}

\begin{document}

\begin{center}
    {\Large\textbf{Relativistic Elastic Scattering of Cosmic-Ray Boosted Dirac Dark Matter with Charged Leptons through a Scalar Mediator}}\\[8pt]
    {\normalsize\textbf{Nguyen Nhu Xuan$^1$, Vu Toan Thang$^2$}}\\[3pt]
    {\footnotesize\textit{$^1$Faculty of Physics and Chemical Engineering, Le Quy Don Technical University,\\
    236 Hoang Quoc Viet Street, Nghia Do Ward, Hanoi, Vietnam\\
    $^2$Faculty of Basic Science, Ngo Quyen University, 229B Bach Dang Street, Thu Dau Mot Ward, Ho Chi Minh City, Vietnam}\\
    \texttt{Xuannn@lqdtu.edu.vn} \quad|\quad ORCID: \href{https://orcid.org/0000-0001-7324-9972}{0000-0001-7324-9972}}
\end{center}

\vspace{2pt}
\begin{abstract}
\noindent
We investigate the relativistic kinematics, spinor-recoil structure, and direct-detection phenomenology of a Dirac dark matter fermion ($\chi$) interacting with Standard Model charged leptons ($\ell$) through a real scalar mediator ($S$) within a low-energy effective Lagrangian in the fermion mass basis. Standard halo direct-detection searches lose sensitivity for sub-MeV dark matter ($m_\chi \lesssim 1\,\mathrm{MeV}$) because the maximum recoil energy falls below detector thresholds. When halo dark matter particles are upscattered to relativistic energies by galactic cosmic-ray electrons (CRE), the momentum transfer in both the astrophysical upscattering stage and the terrestrial detection stage can exceed the fermion mass scales. By decomposing the tree-level $t$-channel cross section into three positive-definite contributions, $\sigma_{\chi\ell}(T_\chi) = \sigma_0 + \sigma_1 + \sigma_2$, we separate the static-numerator contribution ($\sigma_0$, obtained from the constant $16m_\chi^2 m_\ell^2$ trace term) from the linear ($\sigma_1$) and quadratic ($\sigma_2$) momentum-transfer spinor-recoil contributions, evaluated via convergent power series that reduce subtractive cancellation at small momentum transfer. For incident kinetic energies $T_\chi \sim m_\chi$, the spinor-recoil contributions become comparable to and begin to exceed $\sigma_0$; at substantially higher energies ($T_\chi \gg m_\chi$), they enhance the total cross section by several orders of magnitude relative to the static-numerator approximation. Furthermore, retaining the dynamical scalar propagator yields a high-energy cross section that decreases as $1/s$ and satisfies $s$-wave helicity partial-wave unitarity, in contrast to the linear $\mathcal{O}(s/m_S^4)$ growth of the contact approximation. Applying the exact differential recoil spectra within an effective propagation model for cosmic-ray electron boosted dark matter (CRE-BDM), we derive $90\%$ C.L. exclusion bounds in the $\mathrm{keV}$--$10\,\mathrm{MeV}$ mass range from Super-Kamiokande, XENONnT, and PandaX-4T, extending experimental sensitivity into the sub-MeV window ($1\,\mathrm{keV} \le m_\chi < 1\,\mathrm{MeV}$).

\vspace{4pt}

\end{abstract}

\vspace{4pt}
\hrule
\vspace{6pt}

\section{Introduction}\label{sec:intro}
Cosmological and astrophysical observations of galactic rotation curves, gravitational lensing, cosmic microwave background anisotropies, and large-scale structure indicate that non-baryonic dark matter (DM) constitutes approximately $26.5\%$ of the energy density of the Universe~\cite{Planck2018,PDG2024,Bertone,Arcadi}. While Weakly Interacting Massive Particles (WIMPs) with GeV--TeV masses have been extensively searched for via nuclear recoils~\cite{XENONnT2023,LZ2023,PandaX4T2021}, stringent exclusion limits have motivated theoretical and experimental studies of lighter dark matter candidates~\cite{Essig2012,Essig2016}. In \emph{leptophilic dark matter} scenarios, where the dark sector couples predominantly to Standard Model (SM) charged leptons ($\ell = e, \mu, \tau$) at tree level~\cite{Kopp2009,Kopp2014,DEramo2016}, direct nuclear recoils are suppressed and experimental searches rely on low-threshold electron-recoil and ionization detectors such as SENSEI~\cite{SENSEI2018,SENSEI2020}, DAMIC~\cite{DAMIC2019}, SuperCDMS~\cite{SuperCDMS2021}, XENON10~\cite{XENON10_2012}, and XENON1T~\cite{XENON1T_S2_2019}.

Conventional direct-detection searches for virialized dark matter in the Milky Way halo ($v_{\mathrm{DM}} \sim 10^{-3}c$, Lorentz factor $\gamma_{\mathrm{DM}} - 1 \simeq v_{\mathrm{DM}}^2/2 \sim 5 \times 10^{-7}$) are limited by a kinematic threshold. A halo particle of mass $m_\chi$ carries a maximum kinetic energy $T_\chi^{\mathrm{halo}} = \frac{1}{2}m_\chi v_{\mathrm{esc}}^2 \lesssim 3 \times 10^{-6}\,m_\chi$ (for galactic escape speed $v_{\mathrm{esc}} \approx 544\,\mathrm{km/s}$). For $m_\chi \lesssim 1\,\mathrm{MeV}$, the maximum recoil energy transferable to a bound electron drops below the ionization or band-gap threshold ($E_{\mathrm{th}} \sim 1\text{--}10\,\mathrm{eV}$) of terrestrial detectors.

A physical mechanism that provides experimental access to this sub-MeV mass region is \emph{Cosmic-Ray Boosted Dark Matter (CRDM)}~\cite{Bringmann2019,Ema2019,Cappiello2019,Dent2020,Bardhan2023,Wang2022}. Because the galactic halo contains energetic cosmic-ray electrons (CRE) with kinetic energies from MeV to TeV scales~\cite{Boschini2018}, elastic collisions $e_{\mathrm{CR}}^- + \chi \to e_{\mathrm{CR}}^- + \chi$ upscatter a fraction of ambient dark matter particles to relativistic energies ($\gamma_{\mathrm{DM}} \gg 1$). When this cosmic-ray electron boosted dark matter (CRE-BDM) flux reaches terrestrial detectors such as Super-Kamiokande~\cite{SuperK_BDM2023}, XENONnT~\cite{XENONnT2023}, or PandaX-4T~\cite{PandaX4T_BDM2022}, it produces keV-to-GeV electron recoils even for dark matter masses in the sub-MeV range ($1\,\mathrm{keV} \le m_\chi < 1\,\mathrm{MeV}$).

In phenomenological studies of boosted dark matter, the microscopic scattering cross section is sometimes approximated either by keeping only the constant threshold term $16m_\chi^2 m_e^2$ in the spin-averaged Dirac trace numerator $(4m_\chi^2 - t)(4m_e^2 - t)$ (referred to hereafter as the \emph{static-numerator approximation}), or by using a dimension-six Effective Field Theory (EFT) contact operator $\mathcal{L}_{\mathrm{EFT}} = (g_\chi g_e/m_S^2)(\bar{\chi}\chi)(\bar{e}e)$. In relativistic kinematics, both approximations have important limitations:
\begin{enumerate}\setlength{\itemsep}{1pt}\setlength{\parskip}{0pt}
    \item In detector scattering with electron recoil kinetic energy $T_e$, the squared momentum transfer satisfies $|t| = 2m_e T_e$. Thus, the electron-mass scale $|t| \gtrsim 4m_e^2$ is crossed when $T_e \gtrsim 2m_e \simeq 1.02\,\mathrm{MeV}$, whereas probing the dark-matter mass scale $|t| \gtrsim 4m_\chi^2$ requires $T_e \gtrsim 2m_\chi^2 / m_e$. Similarly, in cosmic-ray upscattering with dark matter recoil kinetic energy $T_\chi$, where $|t| = 2m_\chi T_\chi$, crossing $|t| \gtrsim 4m_\chi^2$ requires $T_\chi \gtrsim 2m_\chi$, while crossing $|t| \gtrsim 4m_e^2$ requires $T_\chi \gtrsim 2m_e^2 / m_\chi$. Once these kinematic thresholds are exceeded, the momentum-transfer-dependent spinor terms $4(m_\chi^2 + m_e^2)|t| + t^2$ dominate over the static numerator $16m_\chi^2 m_e^2$.
    \item Conversely, replacing the $t$-channel propagator $(t - m_S^2)^{-2}$ by $m_S^{-4}$ in the EFT contact approximation leads to an unphysical high-energy growth $\sigma_{\mathrm{EFT}}(s) \propto s/m_S^4$ that violates partial-wave unitarity when $\sqrt{s}$ exceeds the mediator mass scale.
\end{enumerate}

Building upon our companion study of the non-relativistic halo regime~\cite{Xuan2026_CIP}, the present paper develops a consistent analytical and numerical treatment of relativistic elastic $\chi + \ell \to \chi + \ell$ scattering and applies it to CRE-BDM phenomenology. Section~\ref{sec:formalism} defines the low-energy effective Lagrangian and derives the exact two-stage relativistic kinematics for upscattering and detection. Section~\ref{sec:decomposition} presents the differential recoil spectra, the decomposition into positive-definite spinor-recoil channels, and its multi-regime numerical validation. Section~\ref{sec:unitarity} analyzes the high-energy asymptotic behavior and $s$-wave helicity partial-wave unitarity. Section~\ref{sec:pheno} computes the CRE-BDM flux, documents the experimental inputs and statistical procedure, and derives $90\%$ C.L. exclusion bounds in the $\mathrm{keV}$--$10\,\mathrm{MeV}$ mass range. Section~\ref{sec:conclusions} summarizes our conclusions.

\section{Effective Model and Relativistic Two-Stage Kinematics}\label{sec:formalism}
\subsection{Low-Energy Effective Lagrangian and Model Assumptions}
We consider a leptophilic simplified model in which the Standard Model is augmented by a gauge-singlet Dirac fermion $\chi \sim (\mathbf{1}, \mathbf{1}, 0)$ and a real scalar mediator $S \sim (\mathbf{1}, \mathbf{1}, 0)$~\cite{Alwall2009,Alves2012,Abercrombie2020}. An exact discrete $\mathbb{Z}_2$ symmetry ($\chi \to -\chi$, $S \to +S$, $\psi_{\mathrm{SM}} \to +\psi_{\mathrm{SM}}$) forbids operators linear in $\chi$ and ensures the cosmological stability of the dark matter candidate.

Above the electroweak symmetry breaking (EWSB) scale, left-handed lepton doublets $L_{\ell L} \sim (\mathbf{1}, \mathbf{2}, -1/2)$ and right-handed charged-lepton singlets $\ell_R \sim (\mathbf{1}, \mathbf{1}, -1)$ transform differently under $\mathrm{SU}(2)_L \times \mathrm{U}(1)_Y$. A gauge-invariant interaction between the singlet scalar $S$ and charged leptons can be generated above the electroweak scale from the dimension-five operator\cite{Kopp2014,DEramo2016}
\begin{equation}
    \mathcal{L}_{\mathrm{UV}}^{(5)} = -\frac{c_\ell}{\Lambda_{\mathrm{UV}}}\,S\bigl(\bar{L}_{\ell L} H \ell_R + \mathrm{h.c.}\bigr),
    \label{eq:L_UV}
\end{equation}
where $H = (0, (v_{\mathrm{EW}} + h)/\sqrt{2})^T$ is the SM Higgs doublet ($v_{\mathrm{EW}} \simeq 246\,\mathrm{GeV}$) and $c_\ell$ is a dimensionless Wilson coefficient, or alternatively through renormalizable portal mixing between $S$ and the Higgs field. After EWSB, Eq.~\eqref{eq:L_UV} yields the effective Yukawa coupling $g_\ell = c_\ell v_{\mathrm{EW}}/(\sqrt{2}\Lambda_{\mathrm{UV}})$ in the fermion mass basis. At momentum transfers well below the electroweak scale ($q = \sqrt{-t} \ll v_{\mathrm{EW}}$), the tree-level scattering dynamics is governed by the low-energy effective Lagrangian\cite{PeskinSchroeder,Schwartz}
\begin{equation}
    \mathcal{L} = \mathcal{L}_{\mathrm{SM}} + \bar{\chi}\bigl(i\slashed{\partial} - m_\chi\bigr)\chi + \frac{1}{2}(\partial_\mu S)(\partial^\mu S) - \frac{1}{2}m_S^2 S^2 - g_\chi S\,\bar{\chi}\chi - g_\ell S\,\bar{\ell}\ell.
    \label{eq:Lagrangian}
\end{equation}
Equation~\eqref{eq:Lagrangian} is the low-energy effective Lagrangian relevant to the tree-level $t$-channel scattering processes studied here, with $m_\chi$ and $m_S$ denoting the physical renormalized masses. Other scalar-potential terms ($S, S^3, S^4$) and Higgs-portal interactions ($S H^\dagger H, S^2 H^\dagger H$) allowed by $\mathbb{Z}_2$ do not contribute to $\chi\ell \to \chi\ell$ at tree level and are omitted in the present calculation; their omission is a simplifying assumption and does not, by itself, define a complete ultraviolet model. In our phenomenological analysis of CRE-BDM ($\ell = e$), we treat $g_\chi$ and $g_e \equiv g_{\ell=e}$ as independent effective parameters; couplings to muons and taus ($g_\mu, g_\tau$) do not enter tree-level dark matter--electron scattering.

For elastic scattering $\chi(p_1, s_1) + \ell(p_2, s_2) \to \chi(p_3, s_3) + \ell(p_4, s_4)$ mediated by $t$-channel scalar exchange, the tree-level invariant amplitude is
\begin{equation}
    i\mathcal{M} = \bigl[\bar{u}_\chi(p_3, s_3)(-ig_\chi)u_\chi(p_1, s_1)\bigr]\,\frac{i}{t - m_S^2}\,\bigl[\bar{u}_\ell(p_4, s_4)(-ig_\ell)u_\ell(p_2, s_2)\bigr],
    \label{eq:M_tree}
\end{equation}
where $s = (p_1 + p_2)^2$, $t = (p_1 - p_3)^2 = (p_2 - p_4)^2 \le 0$, and $u = (p_1 - p_4)^2$ satisfy $s + t + u = 2m_\chi^2 + 2m_\ell^2$. Averaging over initial spins and summing over final spins via $\sum_s u(p,s)\bar{u}(p,s) = \slashed{p} + m$ yields the unpolarized squared matrix element~\cite{Xuan2026_CIP}:
\begin{equation}
    \overline{|\mathcal{M}|^2} = \frac{1}{4}\sum_{\mathrm{spins}}|\mathcal{M}|^2 = \frac{g_\chi^2 g_\ell^2\,(4m_\chi^2 - t)(4m_\ell^2 - t)}{(t - m_S^2)^2}.
    \label{eq:Msq}
\end{equation}

\subsection{Exact Relativistic Kinematics for Upscattering and Detection}
The CRE-BDM process consists of two sequential relativistic two-body collisions in the laboratory frame:
\begin{enumerate}\setlength{\itemsep}{2pt}\setlength{\parskip}{0pt}
    \item \emph{Stage I --- Galactic cosmic-ray upscattering ($e_{\mathrm{CR}}(p_e) + \chi(k_1) \to e_{\mathrm{CR}}(p_e') + \chi(k_2)$):} A cosmic-ray electron with kinetic energy $T_e^{\mathrm{CR}} = E_e^{\mathrm{CR}} - m_e$ and four-momentum $p_e = (E_e^{\mathrm{CR}}, \vec{p}_e)$ scatters off a stationary halo dark matter particle $k_1 = (m_\chi, \vec{0})$ ($v_{\mathrm{halo}} \sim 10^{-3} \ll v_e^{\mathrm{CR}} \simeq 1$), transferring kinetic energy $T_\chi = (\gamma_{\mathrm{DM}} - 1)m_\chi \ge 0$ so that $k_2 = (m_\chi + T_\chi, \vec{k}_2)$. The Mandelstam invariant $s_{\mathrm{CR}} = (p_e + k_1)^2$, the K\"all\'en function $\lambda(s_{\mathrm{CR}}, m_e^2, m_\chi^2)$, and the squared momentum transfer $t = (k_1 - k_2)^2$ are
    \begin{equation}
        s_{\mathrm{CR}} = (m_\chi + m_e)^2 + 2m_\chi T_e^{\mathrm{CR}}, \qquad
        \lambda(s_{\mathrm{CR}}, m_e^2, m_\chi^2) = 4m_\chi^2\,T_e^{\mathrm{CR}}\bigl(T_e^{\mathrm{CR}} + 2m_e\bigr), \qquad
        t = -2m_\chi T_\chi \le 0.
        \label{eq:s_lam_CR}
    \end{equation}
    From the kinematic boundary $|t| \le |t_{\min}^{\mathrm{CR}}| = \lambda(s_{\mathrm{CR}}, m_e^2, m_\chi^2)/s_{\mathrm{CR}}$, the maximum kinetic energy imparted to the dark matter particle and the inverse threshold $T_e^{\mathrm{CR},\min}(T_\chi)$ required to produce a given $T_\chi$ are
    \begin{equation}
        T_\chi^{\max}\bigl(T_e^{\mathrm{CR}}\bigr) = \frac{2m_\chi\,T_e^{\mathrm{CR}}\bigl(T_e^{\mathrm{CR}} + 2m_e\bigr)}{(m_\chi + m_e)^2 + 2m_\chi T_e^{\mathrm{CR}}}, \qquad
        T_e^{\mathrm{CR},\min}(T_\chi) = \left(\frac{T_\chi}{2} - m_e\right) + \sqrt{\frac{T_\chi}{2}\left(\frac{T_\chi}{2} + m_\chi\right)\left(1 + \frac{2m_e^2}{m_\chi T_\chi}\right)}.
        \label{eq:Tchi_max_Te_min}
    \end{equation}

    \item \emph{Stage II --- Terrestrial detector scattering ($\chi(p_1) + e(p_2) \to \chi(p_3) + e(p_4)$):} When the boosted dark matter particle with kinetic energy $T_\chi = E_\chi - m_\chi$ strikes a stationary target electron $p_2 = (m_e, \vec{0})$ in a detector, the electron recoils with kinetic energy $T_e = E_4 - m_e \ge 0$. Interchanging $m_\chi \leftrightarrow m_e$ and $T_e^{\mathrm{CR}} \leftrightarrow T_\chi$, we obtain
    \begin{equation}
        s = (m_\chi + m_e)^2 + 2m_e T_\chi, \qquad
        \lambda(s, m_\chi^2, m_e^2) = 4m_e^2\,T_\chi\bigl(T_\chi + 2m_\chi\bigr), \qquad
        t = -2m_e T_e \le 0,
        \label{eq:s_lam_det}
    \end{equation}
    together with the maximum detector electron recoil energy $T_e^{\max}(T_\chi)$ and the minimum incident dark matter kinetic energy $T_\chi^{\min}(T_e)$ needed to induce a recoil $T_e$:
    \begin{equation}
        T_e^{\max}(T_\chi) = \frac{2m_e\,T_\chi(T_\chi + 2m_\chi)}{(m_\chi + m_e)^2 + 2m_e T_\chi}, \qquad
        T_\chi^{\min}(T_e) = \left(\frac{T_e}{2} - m_\chi\right) + \sqrt{\frac{T_e}{2}\left(\frac{T_e}{2} + m_e\right)\left(1 + \frac{2m_\chi^2}{m_e T_e}\right)}.
        \label{eq:Te_max_Tchi_min}
    \end{equation}
\end{enumerate}

\section{Differential Recoil Spectra, Spinor Decomposition, and Numerical Validation}\label{sec:decomposition}
\subsection{Differential Recoil Cross Sections for Upscattering and Detection}
Using $d\sigma/dt = \overline{|\mathcal{M}|^2}/[16\pi\lambda(s, m_1^2, m_2^2)]$ together with $dt = -2m_\chi\,dT_\chi$ in Stage I and $dt = -2m_e\,dT_e$ in Stage II, the exact tree-level differential recoil spectra are
\begin{align}
    \frac{d\sigma_{\mathrm{CR}}}{dT_\chi}\bigl(T_e^{\mathrm{CR}}, T_\chi\bigr) &= \frac{g_\chi^2 g_e^2}{32\pi\,m_\chi\,T_e^{\mathrm{CR}}\bigl(T_e^{\mathrm{CR}} + 2m_e\bigr)}\,\frac{\bigl(4m_\chi^2 + 2m_\chi T_\chi\bigr)\bigl(4m_e^2 + 2m_\chi T_\chi\bigr)}{\bigl(2m_\chi T_\chi + m_S^2\bigr)^2}\,\Theta\bigl(T_\chi^{\max}(T_e^{\mathrm{CR}}) - T_\chi\bigr), \label{eq:dsigma_CR} \\[3pt]
    \frac{d\sigma_{\chi e}}{dT_e}\bigl(T_\chi, T_e\bigr) &= \frac{g_\chi^2 g_e^2}{32\pi\,m_e\,T_\chi\bigl(T_\chi + 2m_\chi\bigr)}\,\frac{\bigl(4m_\chi^2 + 2m_e T_e\bigr)\bigl(4m_e^2 + 2m_e T_e\bigr)}{\bigl(2m_e T_e + m_S^2\bigr)^2}\,\Theta\bigl(T_e^{\max}(T_\chi) - T_e\bigr). \label{eq:dsigma_det}
\end{align}
To determine when the recoil spectrum $d\sigma_{\chi e}/dT_e$ increases with recoil energy $T_e$, we differentiate the momentum-transfer factor $f(q^2) \equiv (4m_\chi^2 + q^2)(4m_e^2 + q^2)/(q^2 + m_S^2)^2$ with respect to $q^2 = 2m_e T_e > 0$:
\begin{equation}
    \frac{df(q^2)}{dq^2} = \frac{2\Bigl[\bigl(2m_\chi^2 + 2m_e^2 + m_S^2\bigr)q^2 + 2\bigl(m_\chi^2 + m_e^2\bigr)m_S^2 - 16m_\chi^2 m_e^2\Bigr]}{\bigl(q^2 + m_S^2\bigr)^3}.
    \label{eq:df_dq2}
\end{equation}
Equation~\eqref{eq:df_dq2} shows that whereas the static-numerator approximation $16m_\chi^2 m_e^2/(q^2 + m_S^2)^2$ decreases monotonically with $q^2$ for all parameter choices, the exact relativistic spectrum satisfies $df/dq^2 > 0$ for all $q^2 > 0$ whenever $m_S^2 > 8m_\chi^2 m_e^2/(m_\chi^2 + m_e^2)$, and in general increases with $T_e$ over the kinematic interval $q^2 > [16m_\chi^2 m_e^2 - 2(m_\chi^2 + m_e^2)m_S^2]/(2m_\chi^2 + 2m_e^2 + m_S^2)$ up to the endpoint $T_e^{\max}(T_\chi)$.

\subsection{Spinor-Recoil Decomposition and Multi-Regime Numerical Validation}
Expanding the Dirac trace numerator in powers of $q^2 = -t \in [0, |t_{\min}|]$ (with $|t_{\min}| = 2m_\ell T_\ell^{\max} = q_{\max}^2$),
\begin{equation}
    (4m_\chi^2 + q^2)(4m_\ell^2 + q^2) = \underbrace{16m_\chi^2 m_\ell^2}_{\text{Static-numerator}} + \underbrace{4(m_\chi^2 + m_\ell^2)q^2}_{\text{Linear spinor recoil}} + \underbrace{q^4}_{\text{Quadratic spinor recoil}},
    \label{eq:trace_expand}
\end{equation}
and integrating over $0 \le T_\ell \le T_\ell^{\max}$, the total elastic cross section $\sigma_{\chi\ell}(T_\chi)$ decomposes into three positive contributions:
\begin{equation}
    \sigma_{\chi\ell}(T_\chi) = \sigma_0(T_\chi) + \sigma_1(T_\chi) + \sigma_2(T_\chi) = \sigma_{\chi\ell}(T_\chi)\bigl[\mathcal{F}_0(T_\chi) + \mathcal{F}_1(T_\chi) + \mathcal{F}_2(T_\chi)\bigr],
    \label{eq:sigma_decomp}
\end{equation}
where $\mathcal{F}_i(T_\chi) \equiv \sigma_i(T_\chi)/\sigma_{\chi\ell}(T_\chi) \in [0, 1]$ ($\mathcal{F}_0 + \mathcal{F}_1 + \mathcal{F}_2 = 1$). In terms of the dimensionless ratio $x \equiv |t_{\min}|/m_S^2 = 2m_\ell T_\ell^{\max}(T_\chi)/m_S^2$, the three partial cross sections are
\begin{align}
    \sigma_0(T_\chi) &= \frac{g_\chi^2 g_\ell^2}{16\pi\,\lambda(s, m_\chi^2, m_\ell^2)}\,\frac{16m_\chi^2 m_\ell^2\,|t_{\min}|}{m_S^2(m_S^2 + |t_{\min}|)} = \frac{g_\chi^2 g_\ell^2\,m_\chi^2 m_\ell^2}{\pi\,\lambda(s, m_\chi^2, m_\ell^2)\,m_S^2}\,\frac{x}{1 + x}, \label{eq:sigma0} \\[2pt]
    \sigma_1(T_\chi) &= \frac{g_\chi^2 g_\ell^2\,(m_\chi^2 + m_\ell^2)}{4\pi\,\lambda(s, m_\chi^2, m_\ell^2)}\,\Phi_1(x), \qquad\quad
    \sigma_2(T_\chi) = \frac{g_\chi^2 g_\ell^2\,m_S^2}{16\pi\,\lambda(s, m_\chi^2, m_\ell^2)}\,\Phi_2(x). \label{eq:sigma12}
\end{align}
Here $\sigma_0(T_\chi)$ is the \emph{static-numerator contribution} (or \emph{static-numerator approximation}), obtained by keeping only the constant $16m_\chi^2 m_\ell^2$ term in Eq.~\eqref{eq:trace_expand} while retaining the exact relativistic phase space and propagator. The auxiliary recoil integrals $\Phi_1(x)$ and $\Phi_2(x)$ and their exact Maclaurin and $(1+x)^{-1}$-factored series representations are
\begin{equation}
\begin{aligned}
    \Phi_1(x) &\equiv \ln(1 + x) - \frac{x}{1 + x} = \sum_{n=0}^{\infty}(-1)^n\frac{n+1}{n+2} x^{n+2} = \frac{x^2}{1 + x}\sum_{n=0}^{\infty}\frac{(-x)^n}{(n + 1)(n + 2)}, \\
    \Phi_2(x) &\equiv x - 2\ln(1 + x) + \frac{x}{1 + x} = \sum_{n=0}^{\infty}(-1)^n\frac{n+1}{n+3} x^{n+3} = \frac{2x^3}{1 + x}\sum_{n=0}^{\infty}\frac{(-x)^n}{(n + 2)(n + 3)}.
\end{aligned}
\label{eq:Phi_series}
\end{equation}

The reformulated expressions in Eqs.~\eqref{eq:sigma_decomp}--\eqref{eq:Phi_series} are designed to reduce subtractive cancellation in the small-$x$ regime. Specifically, while direct floating-point evaluation of the closed-form definitions of $\Phi_1(x)$ and $\Phi_2(x)$ suffers from internal cancellation of $\mathcal{O}(x)$ and $\mathcal{O}(x^2)$ terms when $x \ll 1$, summing the factored series in Eq.~\eqref{eq:Phi_series} for $0 \le x < x_{\mathrm{th}} = 0.5$ to $N_{\mathrm{trunc}} = 50$ terms bounds the truncation remainder below $2x_{\mathrm{th}}^{53}/(52\times 53) \simeq 8 \times 10^{-20} \ll \epsilon_{\mathrm{mach}} = 2^{-53} \simeq 1.11 \times 10^{-16}$. Conversely, for $x \ge 0.5$ (where $\Phi_1(x) \ge 0.072$ and $\Phi_2(x) \ge 0.022$), evaluating the closed-form expressions with $\mathrm{log1p}(x)$ loses at most one decimal digit. Table~\ref{tab:validation} assesses the numerical accuracy of this double-precision scheme ($\sigma_{\mathrm{stable}}$) and of naive antiderivative subtraction ($\sigma_{\mathrm{naive}}$) against 80-digit arbitrary-precision reference calculations ($\sigma_{\mathrm{ref}}$, cross-checked with adaptive Gauss--Kronrod quadrature) across representative mass hierarchies ($m_\chi \ll m_e$, $m_\chi \sim m_e$, $m_\chi \gg m_e$) and kinematic regimes ($x \ll 1$, $x \sim 1$, $x \gg 1$), confirming a relative accuracy $\varepsilon_{\mathrm{stable}} \equiv |\sigma_{\mathrm{stable}} - \sigma_{\mathrm{ref}}|/\sigma_{\mathrm{ref}} \le 1.2 \times 10^{-15}$ over the tested range.

\begin{table}[htbp]
    \centering
    \caption{Multi-regime numerical validation of the double-precision reformulation $\sigma_{\mathrm{stable}}$ [Eqs.~\eqref{eq:sigma_decomp}--\eqref{eq:Phi_series}] and naive antiderivative subtraction $\sigma_{\mathrm{naive}}$ against 80-digit reference calculations $\sigma_{\mathrm{ref}}$ across mass hierarchies ($m_\chi \ll m_e$, $m_\chi \sim m_e$, $m_\chi \gg m_e$) and momentum-transfer regimes ($x \ll 1$, $x \sim 1$, $x \gg 1$) for $g_\chi = g_e = 0.1$. Here $\varepsilon_{\mathrm{stable}} \equiv |\sigma_{\mathrm{stable}} - \sigma_{\mathrm{ref}}|/\sigma_{\mathrm{ref}}$ and $\varepsilon_{\mathrm{naive}} \equiv |\sigma_{\mathrm{naive}} - \sigma_{\mathrm{ref}}|/\sigma_{\mathrm{ref}}$.}
    \label{tab:validation}
    \resizebox{\textwidth}{!}{%
    \begin{tabular}{lccccccc}
        \toprule
        Kinematic Regime & $m_\chi\,[\mathrm{GeV}]$ & $T_\chi\,[\mathrm{GeV}]$ & $m_S\,[\mathrm{GeV}]$ & $x = |t_{\min}|/m_S^2$ & $\sigma_{\mathrm{ref}}\,[\mathrm{cm^2}]$ & $\varepsilon_{\mathrm{stable}}$ & $\varepsilon_{\mathrm{naive}}$ \\
        \midrule
        $m_\chi \ll m_e,\; x \ll 1$ & $10^{-5}$        & $10^{-6}$ & $10^{-2}$        & $8.05 \times 10^{-7}$ & $1.3074 \times 10^{-34}$ & $6.5 \times 10^{-16}$ & $5.4 \times 10^{-2}$ \\
        $m_\chi \ll m_e,\; x \gg 1$ & $10^{-5}$        & $10^{-2}$ & $10^{-3}$        & $9.98 \times 10^{0}$  & $5.6601 \times 10^{-29}$ & $5.9 \times 10^{-16}$ & $2.0 \times 10^{-16}$ \\
        $m_\chi \sim m_e,\; x \ll 1$ & $5\times 10^{-4}$ & $10^{-4}$ & $10^{-2}$        & $1.02 \times 10^{-3}$ & $7.9317 \times 10^{-32}$ & $5.5 \times 10^{-16}$ & $1.0 \times 10^{-8}$ \\
        $m_\chi \sim m_e,\; x \sim 1$ & $5\times 10^{-4}$ & $5\times 10^{-2}$ & $5\times 10^{-3}$ & $2.04 \times 10^{0}$  & $3.8274 \times 10^{-30}$ & $1.8 \times 10^{-16}$ & $2.2 \times 10^{-15}$ \\
        $m_\chi \gg m_e,\; x \ll 1$ & $10^{-2}$        & $10^{-6}$ & $10^{-2}$        & $1.89 \times 10^{-6}$ & $2.9296 \times 10^{-31}$ & $7.5 \times 10^{-16}$ & $5.3 \times 10^{-8}$ \\
        $m_\chi \gg m_e,\; x \sim 1$ & $10^{-2}$        & $10^{-1}$ & $10^{-2}$        & $5.89 \times 10^{-1}$ & $2.5958 \times 10^{-30}$ & $1.2 \times 10^{-15}$ & $4.9 \times 10^{-15}$ \\
        $m_\chi \gg m_e,\; x \gg 1$ & $10^{-2}$        & $10^{1}$  & $10^{-2}$        & $1.01 \times 10^{2}$  & $7.9698 \times 10^{-32}$ & $8.2 \times 10^{-16}$ & $4.1 \times 10^{-16}$ \\
        $m_\chi \gg m_e,\; x \gg 1$ & $1.0$            & $10^{2}$  & $10^{-2}$        & $9.66 \times 10^{1}$  & $1.0480 \times 10^{-30}$ & $1.7 \times 10^{-16}$ & $1.7 \times 10^{-16}$ \\
        \bottomrule
    \end{tabular}%
    }
\end{table}

\section{High-Energy Asymptotics and Partial-Wave Unitarity}\label{sec:unitarity}
\subsection{High-Energy Asymptotic Scaling Relative to the Static-Numerator Approximation}
In the ultra-relativistic limit $s \simeq 2m_\ell T_\chi \gg m_\chi^2, m_\ell^2, m_S^2$, we have $\lambda(s, m_\chi^2, m_\ell^2) \simeq s^2$, $|t_{\min}| \simeq s$, and $x = |t_{\min}|/m_S^2 \simeq s/m_S^2 \gg 1$. Using $\Phi_1(x) \simeq \ln x - 1$ and $\Phi_2(x) \simeq x - 2\ln x$, the three partial cross sections in Eqs.~\eqref{eq:sigma0}--\eqref{eq:sigma12} exhibit the asymptotic forms
\begin{equation}
    \sigma_0(s) \simeq \frac{g_\chi^2 g_\ell^2 m_\chi^2 m_\ell^2}{\pi m_S^2 s^2} \propto \frac{1}{s^2}, \quad\;
    \sigma_1(s) \simeq \frac{g_\chi^2 g_\ell^2(m_\chi^2 + m_\ell^2)}{4\pi s^2}\ln\!\left(\frac{s}{e m_S^2}\right) \propto \frac{\ln s}{s^2}, \quad\;
    \sigma_2(s) \simeq \frac{g_\chi^2 g_\ell^2}{16\pi s}\left[1 - \frac{2m_S^2}{s}\ln\!\left(\frac{s}{m_S^2}\right)\right] \propto \frac{1}{s}.
    \label{eq:asymptotic_s}
\end{equation}
The ratio of the full cross section $\sigma_{\chi\ell}(s)$ to the static-numerator contribution $\sigma_0(s)$ is given for arbitrary $x$ and in the asymptotic regime $s \gg m_\chi^2, m_\ell^2, m_S^2$ by
\begin{equation}
    \frac{\sigma_{\chi\ell}(s)}{\sigma_0(s)} = 1 + \frac{1 + x}{4m_\chi^2 m_\ell^2 x}\Bigl[4(m_\chi^2 + m_\ell^2)\Phi_1(x) + m_S^2 \Phi_2(x)\Bigr] \;\underset{s \gg m_\chi^2, m_\ell^2, m_S^2}{\simeq}\; 1 + \frac{(m_\chi^2 + m_\ell^2)m_S^2}{4m_\chi^2 m_\ell^2}\ln\!\left(\frac{s}{e\,m_S^2}\right) + \frac{m_S^2\,s}{16m_\chi^2 m_\ell^2}.
    \label{eq:enhancement_ratio}
\end{equation}
At large momentum transfer, the $q^4$ term in the spin-averaged numerator~\eqref{eq:trace_expand} controls the high-energy behavior of the scalar-mediated scattering rate ($\mathcal{F}_2 \to 1$), causing $\sigma_{\chi\ell}(s)$ to exceed the static-numerator approximation $\sigma_0(s) \propto s^{-2}$ by a factor proportional to $m_S^2 s / (16m_\chi^2 m_\ell^2)$ for $s \gg m_\chi^2, m_\ell^2, m_S^2$.

\subsection{Helicity Amplitudes and \texorpdfstring{$S$}{S}-Wave Partial-Wave Unitarity}
In the dimension-six EFT contact approximation $\mathcal{L}_{\mathrm{EFT}} = (g_\chi g_\ell / m_S^2)(\bar{\chi}\chi)(\bar{\ell}\ell)$, replacing $(t - m_S^2)^2 \to m_S^4$ yields
\begin{equation}
    \sigma_{\mathrm{EFT}}(s) = \frac{g_\chi^2 g_\ell^2\,|t_{\min}|}{16\pi\,\lambda(s, m_\chi^2, m_\ell^2)\,m_S^4}\left[16m_\chi^2 m_\ell^2 + 2(m_\chi^2 + m_\ell^2)|t_{\min}| + \frac{1}{3}|t_{\min}|^2\right] \;\underset{s \gg m_\chi^2, m_\ell^2}{\simeq}\; \frac{g_\chi^2 g_\ell^2\,s}{48\pi\,m_S^4}.
    \label{eq:sigma_EFT_high}
\end{equation}
To examine how the full scalar propagator cures the linear high-energy growth of Eq.~\eqref{eq:sigma_EFT_high} at the amplitude level, we evaluate the CM-frame helicity amplitudes $\mathcal{M}_{\lambda_3\lambda_4;\lambda_1\lambda_2}(s, \theta)$, where $\lambda_i \in \{+1/2, -1/2\}$ denote the fermion helicities. In the ultra-relativistic regime $s \gg m_\chi^2, m_\ell^2$, the scalar bilinear $\bar{u}(p', \lambda')u(p, \lambda)$ flips fermion helicity at each vertex, $|\bar{u}(p_3, \lambda_3)u(p_1, \lambda_1)| \simeq \sqrt{s}\,\sin(\theta/2)\,\delta_{\lambda_3, -\lambda_1} + \mathcal{O}(m_\chi)$. Consequently, the four helicity-flip channels $(\lambda_3, \lambda_4) = (-\lambda_1, -\lambda_2)$ dominate and have equal magnitude:
\begin{equation}
    \bigl|\mathcal{M}_{-\lambda_1, -\lambda_2; \lambda_1, \lambda_2}(s, \theta)\bigr| \;\underset{s \gg m_\chi^2, m_\ell^2}{\simeq}\; \frac{g_\chi g_\ell\,s\sin^2(\theta/2)}{|t| + m_S^2} = \frac{g_\chi g_\ell\,q^2}{q^2 + m_S^2},
    \label{eq:M_helicity}
\end{equation}
where $q^2 = -t = \frac{s}{2}(1 - \cos\theta) \in [0, s]$. In the Jacob--Wick formalism~\cite{PeskinSchroeder,Schwartz}, the partial-wave expansion of a helicity amplitude with initial net helicity $\lambda = \lambda_1 - \lambda_2$ and final net helicity $\mu = \lambda_3 - \lambda_4$ is $\mathcal{M}_{\lambda_3\lambda_4;\lambda_1\lambda_2}(s, \theta) = 16\pi \sum_J (2J + 1)\,a^J_{\mu\lambda}(s)\,d^J_{\lambda\mu}(\theta)$. For the equal-helicity channels $(\lambda_1, \lambda_2) = (\pm 1/2, \pm 1/2)$, we have $\lambda = \mu = 0$ and $d^0_{00}(\theta) = P_0(\cos\theta) = 1$, so the $J = 0$ ($s$-wave) partial-wave projection is
\begin{equation}
    a^0_{00}(s) = \frac{1}{32\pi}\int_{-1}^{1} d(\cos\theta)\,\mathcal{M}_{\mp\mp;\pm\pm}(s, \theta) \;\underset{s \gg m_\chi^2, m_\ell^2}{\simeq}\; -\frac{g_\chi g_\ell}{16\pi}\int_{0}^{s}\frac{dq^2}{s}\,\frac{q^2}{q^2 + m_S^2} = -\frac{g_\chi g_\ell}{16\pi}\left[1 - \frac{m_S^2}{s}\ln\!\left(1 + \frac{s}{m_S^2}\right)\right].
    \label{eq:a0_exact}
\end{equation}
Under this normalization, elastic $s$-wave unitarity requires $|\mathrm{Re}\,a^0_{00}(s)| \le 1/2$. In the EFT contact approximation ($q^2 + m_S^2 \to m_S^2$), $|a_0^{\mathrm{EFT}}(s)| = g_\chi g_\ell s / (32\pi m_S^2)$ grows linearly with $s$ and violates $|\mathrm{Re}\,a_0^{\mathrm{EFT}}(s)| \le 1/2$ above
\begin{equation}
    \sqrt{s_{\mathrm{unit}}^{\mathrm{EFT}}} = \sqrt{\frac{16\pi}{g_\chi g_\ell}}\,m_S.
    \label{eq:s_unit_EFT}
\end{equation}
When the dynamical scalar propagator is retained in Eq.~\eqref{eq:a0_exact}, $|a^0_{00}(s)|$ approaches the bounded constant $g_\chi g_\ell / (16\pi)$ for $s \gg m_S^2$ (satisfying $|\mathrm{Re}\,a^0_{00}(s)| \le 1/2$ throughout the perturbative regime $g_\chi, g_\ell \lesssim \sqrt{4\pi}$), and the total cross section decreases as $1/s$:
\begin{equation}
    \lim_{s \gg m_\chi^2, m_\ell^2, m_S^2} \sigma_{\chi\ell}(s) = \frac{g_\chi^2 g_\ell^2}{16\pi\,s}.
    \label{eq:unitary_asymptote}
\end{equation}

\section{Numerical Analysis and Cosmic-Ray Boosted Dark Matter Bounds}\label{sec:pheno}
\subsection{Relativistic Cross Section Enhancement and Spinor Weights}
\begin{figure}[htbp]
    \centering
    \includegraphics[width=0.95\textwidth]{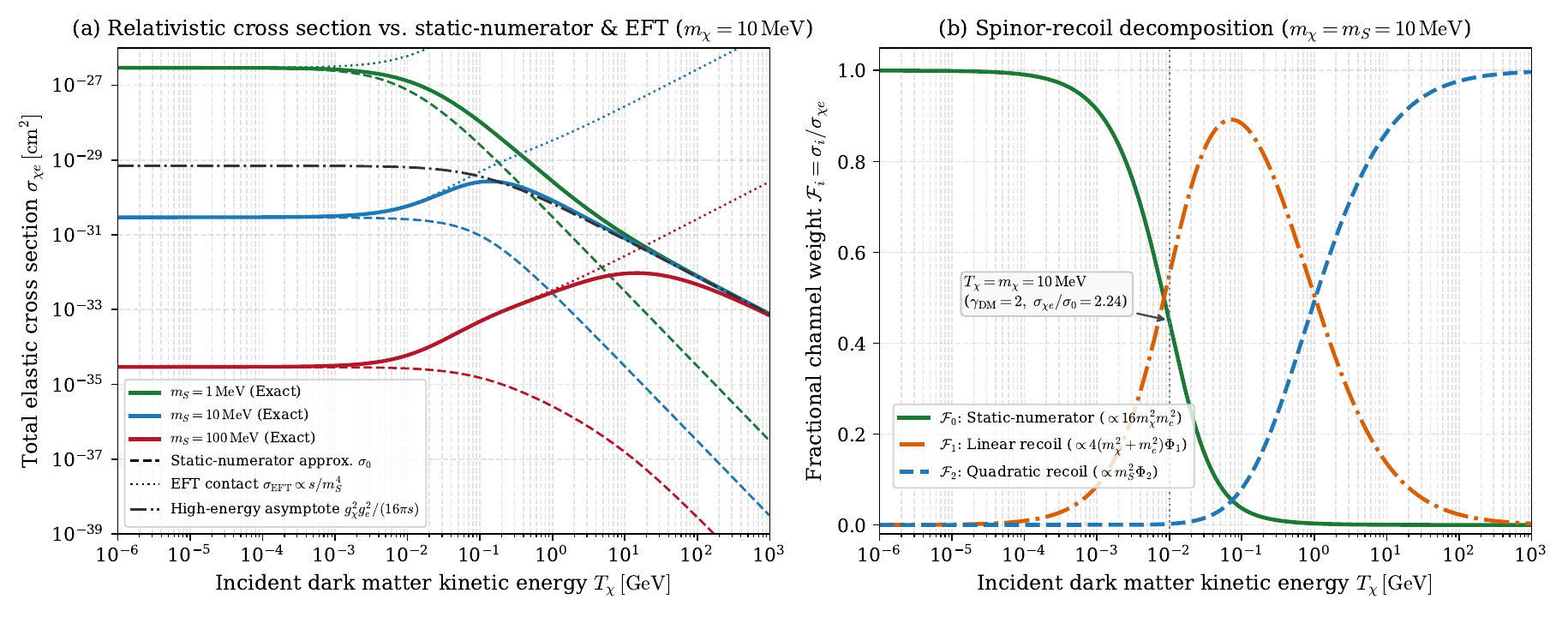}
    \caption{(a) Total elastic cross section $\sigma_{\chi e}(T_\chi)$ versus incident dark matter kinetic energy $T_\chi \in [10^{-6}, 10^{3}]\,\mathrm{GeV}$ for $m_\chi = 10\,\mathrm{MeV}$, $m_S \in \{1, 10, 100\}\,\mathrm{MeV}$, and $g_\chi = g_e = 0.1$, comparing the exact relativistic result (solid curves), the static-numerator approximation $\sigma_0$ (dashed curves), the EFT contact approximation $\sigma_{\mathrm{EFT}}$ (dotted curves), and the high-energy asymptote $g_\chi^2 g_e^2/(16\pi s)$ (dash-dotted line). (b) Fractional channel weights $\mathcal{F}_0$, $\mathcal{F}_1$, and $\mathcal{F}_2$ defined in Eq.~\eqref{eq:sigma_decomp} versus $T_\chi$ for the benchmark $m_\chi = m_S = 10\,\mathrm{MeV}$ and $g_\chi = g_e = 0.1$.}
    \label{fig:mpla_fig1}
\end{figure}

We now evaluate Eqs.~\eqref{eq:sigma_decomp}--\eqref{eq:Phi_series} for dark matter--electron scattering ($\ell = e$, $m_e = 0.51099895\,\mathrm{MeV}$) with perturbative couplings $g_\chi = g_e = 0.1$. Figure~\ref{fig:mpla_fig1}(a) compares the exact total cross section $\sigma_{\chi e}(T_\chi)$ (solid curves), the static-numerator approximation $\sigma_0(T_\chi)$ (dashed curves), and the EFT contact cross section $\sigma_{\mathrm{EFT}}(T_\chi)$ (dotted curves) over $T_\chi \in [10^{-6}, 10^3]\,\mathrm{GeV}$ for $m_\chi = 10\,\mathrm{MeV}$ and $m_S \in \{1, 10, 100\}\,\mathrm{MeV}$.

\begin{table}[htbp]
    \centering
    \caption{Benchmark comparison of the exact cross section $\sigma_{\chi e}(T_\chi)$, the static-numerator approximation $\sigma_0(T_\chi)$, the ratio $\sigma_{\chi e}/\sigma_0$, and the fractional channel weights $(\mathcal{F}_0, \mathcal{F}_1, \mathcal{F}_2)$ across incident kinetic energies $T_\chi$ for $m_\chi = m_S = 10\,\mathrm{MeV}$ and $g_\chi = g_e = 0.1$.}
    \label{tab:mpla_bench}
    \resizebox{\textwidth}{!}{%
    \begin{tabular}{ccccccccc}
        \toprule
        $T_\chi\,[\mathrm{GeV}]$ & $\gamma_{\mathrm{DM}}$ & $T_e^{\max}\,[\mathrm{GeV}]$ & $\sigma_{\chi e}\,[\mathrm{cm^2}]$ & $\sigma_0\,[\mathrm{cm^2}]$ & $\sigma_{\chi e}/\sigma_0$ & $\mathcal{F}_0$ & $\mathcal{F}_1$ & $\mathcal{F}_2$ \\
        \midrule
        $10^{-6}$ & $1.0001$  & $1.850 \times 10^{-7}$ & $2.930 \times 10^{-31}$ & $2.929 \times 10^{-31}$ & $1.000$               & $0.9999$ & $0.0001$ & $0.0000$ \\
        $10^{-3}$ & $1.1000$  & $1.925 \times 10^{-4}$ & $3.170 \times 10^{-31}$ & $2.897 \times 10^{-31}$ & $1.094$               & $0.9138$ & $0.0862$ & $0.0000$ \\
        $10^{-2}$ & $2.0000$  & $2.540 \times 10^{-3}$ & $5.856 \times 10^{-31}$ & $2.613 \times 10^{-31}$ & $2.241$               & $0.4463$ & $0.5513$ & $0.0024$ \\
        $10^{-1}$ & $11.000$  & $5.766 \times 10^{-2}$ & $2.596 \times 10^{-30}$ & $9.575 \times 10^{-32}$ & $2.711 \times 10^{1}$ & $0.0369$ & $0.8833$ & $0.0798$ \\
        $1.0$     & $101.00$  & $9.205 \times 10^{-1}$ & $8.313 \times 10^{-31}$ & $2.746 \times 10^{-33}$ & $3.028 \times 10^{2}$ & $0.0033$ & $0.5046$ & $0.4921$ \\
        $10.0$    & $1001.0$  & $9.913 \times 10^{0}$  & $7.970 \times 10^{-32}$ & $3.062 \times 10^{-35}$ & $2.603 \times 10^{3}$ & $0.0004$ & $0.1355$ & $0.8641$ \\
        $100.0$   & $10001.0$ & $9.991 \times 10^{1}$  & $7.653 \times 10^{-33}$ & $3.095 \times 10^{-37}$ & $2.473 \times 10^{4}$ & $0.0000$ & $0.0230$ & $0.9769$ \\
        \bottomrule
    \end{tabular}%
    }
\end{table}

As shown in Fig.~\ref{fig:mpla_fig1} and Table~\ref{tab:mpla_bench} for the benchmark $m_\chi = m_S = 10\,\mathrm{MeV}$, at non-relativistic energies ($T_\chi \le 10^{-4}\,\mathrm{GeV}$) the static-numerator channel dominates ($\mathcal{F}_0 > 0.99$). When the incident kinetic energy reaches the dark matter mass scale $T_\chi = m_\chi = 10\,\mathrm{MeV}$ ($\gamma_{\mathrm{DM}} = 2$), the spinor-recoil contributions begin to exceed the static-numerator contribution ($\mathcal{F}_1 = 0.5513$, $\mathcal{F}_2 = 0.0024$, $\mathcal{F}_0 = 0.4463$), giving a moderate enhancement $\sigma_{\chi e}/\sigma_0 = 2.241$. At substantially higher energies ($T_\chi \gg m_\chi$), $\sigma_0$ falls as $T_\chi^{-2}$ while $\sigma_1 + \sigma_2$ enhances the total cross section by several orders of magnitude: $\sigma_{\chi e}/\sigma_0 = 27.11$ at $T_\chi = 0.1\,\mathrm{GeV}$, $3.028 \times 10^2$ at $T_\chi = 1\,\mathrm{GeV}$, and $2.473 \times 10^4$ at $T_\chi = 100\,\mathrm{GeV}$ ($\mathcal{F}_2 = 0.9769$), where all exact curves converge to the $1/s$ asymptote~\eqref{eq:unitary_asymptote}.

\subsection{Effective-Propagation Model for the CRE-BDM Flux and Recoil Spectra}
We estimate the galactic CRE-boosted dark matter flux arriving at Earth within the standard effective-propagation model~\cite{Bringmann2019,Ema2019,Cappiello2019,Dent2020}:
\begin{equation}
    \frac{d\Phi_\chi}{dT_\chi} = D_{\mathrm{eff}}\,\frac{\rho_\chi^{\mathrm{local}}}{m_\chi}\int_{T_e^{\mathrm{CR},\min}(T_\chi)}^{\infty} dT_e^{\mathrm{CR}}\,\frac{d\Phi_e^{\mathrm{CR}}}{dT_e^{\mathrm{CR}}}\,\frac{d\sigma_{\mathrm{CR}}\bigl(T_e^{\mathrm{CR}}, T_\chi\bigr)}{dT_\chi}.
    \label{eq:dPhi_chi}
\end{equation}
Here $\rho_\chi^{\mathrm{local}} = 0.3\,\mathrm{GeV/cm^3}$ is the local dark matter density, $T_e^{\mathrm{CR},\min}(T_\chi)$ is given by Eq.~\eqref{eq:Tchi_max_Te_min}, and $d\Phi_e^{\mathrm{CR}}/dT_e^{\mathrm{CR}} = 4\pi\,dI_e^{\mathrm{CR}}/dT_e^{\mathrm{CR}}$ (in units of $\mathrm{cm^{-2}\,s^{-1}\,GeV^{-1}}$) is the angle-integrated omnidirectional local interstellar cosmic-ray electron flux parameterized from Voyager-1 and AMS-02 data~\cite{Boschini2018,Cappiello2019}. The effective propagation distance $D_{\mathrm{eff}} = (4\pi \rho_\chi^{\mathrm{local}})^{-1}\int d\Omega\int_{\mathrm{l.o.s.}} d\ell\,\rho_\chi(r)\,(\Phi_e^{\mathrm{CR}}(r)/\Phi_e^{\mathrm{CR},\mathrm{local}})$ encapsulates the line-of-sight integral over the galactic dark matter halo and cosmic-ray diffusion cylinder; following Refs.~\cite{Cappiello2019,Dent2020}, we adopt $D_{\mathrm{eff}} = 1\,\mathrm{kpc} = 3.086 \times 10^{21}\,\mathrm{cm}$ as a conservative benchmark (whereas full-sky diffusion models with scale height $z_h \sim 4\,\mathrm{kpc}$ yield $D_{\mathrm{eff}} \simeq 8\text{--}10\,\mathrm{kpc}$). The flux in Eq.~\eqref{eq:dPhi_chi} represents the unattenuated relativistic flux at the top of the detector: at the weakly coupled exclusion boundary ($\sqrt{g_\chi g_e} \lesssim 0.8$ for $m_S = 10\,\mathrm{MeV}$, where $\sigma_{\chi e} \lesssim 10^{-31}\,\mathrm{cm^2}$), the dark matter mean free path in rock and the atmosphere exceeds $10^4\,\mathrm{km}$, so overburden attenuation is negligible and is not included.

\begin{figure}[htbp]
    \centering
    \includegraphics[width=0.95\textwidth]{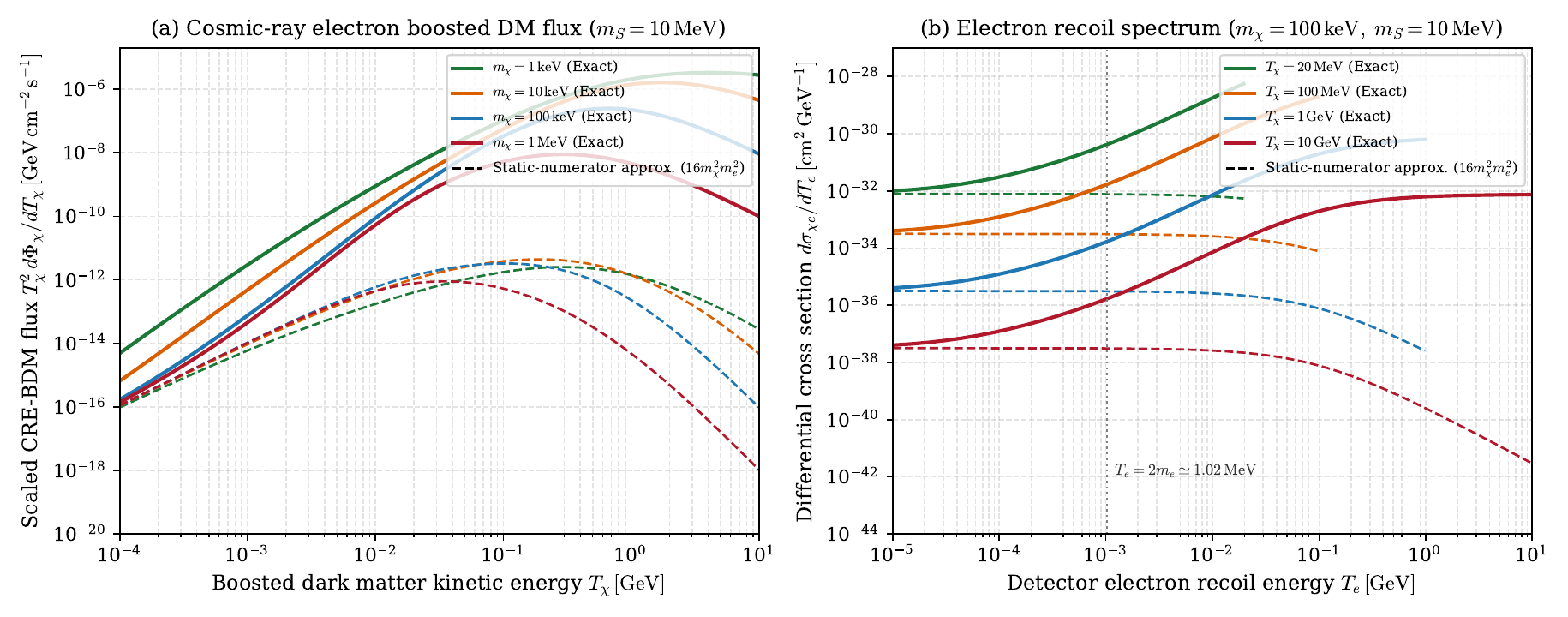}
    \caption{(a) Scaled CRE-BDM flux $T_\chi^2\,d\Phi_\chi/dT_\chi$ estimated within the effective-propagation model ($D_{\mathrm{eff}} = 1\,\mathrm{kpc}$) versus $T_\chi \in [10^{-4}, 10^{1}]\,\mathrm{GeV}$ for $m_\chi = 1\,\mathrm{keV}$, $10\,\mathrm{keV}$, $100\,\mathrm{keV}$, and $1\,\mathrm{MeV}$ at $m_S = 10\,\mathrm{MeV}$ and $g_\chi = g_e = 0.1$, comparing the exact relativistic calculation (solid curves) with the static-numerator approximation (dashed curves). (b) Differential electron recoil cross section $d\sigma_{\chi e}/dT_e$ versus $T_e \in [10^{-5}, 10^{1}]\,\mathrm{GeV}$ for $T_\chi \in \{20\,\mathrm{MeV}, 100\,\mathrm{MeV}, 1\,\mathrm{GeV}, 10\,\mathrm{GeV}\}$ at the benchmark $m_\chi = 100\,\mathrm{keV}$, $m_S = 10\,\mathrm{MeV}$, and $g_\chi = g_e = 0.1$.}
    \label{fig:mpla_fig2}
\end{figure}

Figure~\ref{fig:mpla_fig2}(a) displays the scaled CRE-BDM flux $T_\chi^2\,d\Phi_\chi/dT_\chi$ for $m_\chi = 1\,\mathrm{keV}$, $10\,\mathrm{keV}$, $100\,\mathrm{keV}$, and $1\,\mathrm{MeV}$ at $m_S = 10\,\mathrm{MeV}$ and $g_\chi = g_e = 0.1$. For $T_\chi \gtrsim 10\,\mathrm{MeV}$ (where $|t| = 2m_\chi T_\chi$ exceeds both $4m_\chi^2$ and $4m_e^2$), the exact relativistic flux exceeds the static-numerator approximation by $2$ to $5$ orders of magnitude. Figure~\ref{fig:mpla_fig2}(b) shows the detector recoil spectrum $d\sigma_{\chi e}/dT_e$ [Eq.~\eqref{eq:dsigma_det}] for the benchmark $m_\chi = 100\,\mathrm{keV}$ and $m_S = 10\,\mathrm{MeV}$. Because this benchmark satisfies $m_S^2 \gg 8m_\chi^2 m_e^2/(m_\chi^2 + m_e^2)$ in Eq.~\eqref{eq:df_dq2}, the exact recoil spectrum rises over the recoil-energy interval $T_e \gtrsim 2m_e \simeq 1.02\,\mathrm{MeV}$ up to the endpoint $T_e^{\max}(T_\chi)$, whereas the static-numerator approximation remains flat for $2m_e T_e \ll m_S^2$ and falls as $T_e^{-2}$ for $2m_e T_e \gg m_S^2$.

\subsection{Experimental Inputs, Statistical Procedure, and Exclusion Bounds}
The differential recoil rate per target electron in a terrestrial detector is
\begin{equation}
    \frac{dR_e}{dT_e} = \int_{T_\chi^{\min}(T_e)}^{\infty} dT_\chi\,\frac{d\Phi_\chi}{dT_\chi}\,\frac{d\sigma_{\chi e}(T_\chi, T_e)}{dT_e}.
    \label{eq:dRe_dTe}
\end{equation}
To derive reproducible $90\%$ confidence level (C.L.) exclusion bounds from published data, we compute the total predicted signal events $N_{\mathrm{sig}}$ over each experiment's analysis window $[T_e^{\mathrm{th}}, T_e^{\mathrm{upper}}]$:
\begin{equation}
    N_{\mathrm{sig}}(m_\chi, m_S; g_\chi g_e) = N_e\,\Delta t\,\bar{\epsilon}_{\mathrm{det}}\int_{T_e^{\mathrm{th}}}^{T_e^{\mathrm{upper}}} dT_e\,\frac{dR_e}{dT_e},
    \label{eq:N_sig}
\end{equation}
where $N_e \Delta t$ is the electron exposure and $\bar{\epsilon}_{\mathrm{det}}$ is the average detection efficiency across the energy window. Table~\ref{tab:exp_inputs} summarizes the experimental inputs and $90\%$ C.L. signal event upper limits $N_{\mathrm{sig}}^{90\%}$ adopted for Super-Kamiokande IV~\cite{SuperK_BDM2023}, XENONnT~\cite{XENONnT2023}, and PandaX-4T~\cite{PandaX4T_BDM2022}. Because $d\Phi_\chi/dT_\chi \propto (g_\chi g_e)^2$ and $d\sigma_{\chi e}/dT_e \propto (g_\chi g_e)^2$, the signal prediction scales as $N_{\mathrm{sig}} \propto (g_\chi g_e)^4 = (\sqrt{g_\chi g_e})^8$. Evaluating $N_{\mathrm{sig}}^{\mathrm{ref}}(m_\chi, m_S)$ at a reference coupling $g_{\mathrm{ref}} \equiv (g_\chi g_e)_{\mathrm{ref}} = 10^{-2}$ ($\sqrt{g_{\mathrm{ref}}} = 0.1$) via numerical double integration of Eqs.~\eqref{eq:dPhi_chi}--\eqref{eq:N_sig}, the $90\%$ C.L. upper bound on the effective coupling is obtained directly by solving $N_{\mathrm{sig}} \le N_{\mathrm{sig}}^{90\%}$:
\begin{equation}
    \sqrt{g_\chi g_e}^{\mathrm{max}}(m_\chi, m_S) = \sqrt{g_{\mathrm{ref}}}\left[\frac{N_{\mathrm{sig}}^{90\%}}{N_{\mathrm{sig}}^{\mathrm{ref}}(m_\chi, m_S)}\right]^{1/8}.
    \label{eq:g_solve}
\end{equation}
To display the CRE-BDM constraints alongside standard halo bounds, we map $\sqrt{g_\chi g_e}^{\mathrm{max}}$ onto the conventional reference cross section defined at $q_0 = \alpha m_e \simeq 3.73\,\mathrm{keV}$~\cite{Essig2012,Xuan2026_CIP}:
\begin{equation}
    \bar{\sigma}_e = \frac{g_\chi^2 g_e^2\,\mu_{\chi e}^2}{\pi\,(\alpha^2 m_e^2 + m_S^2)^2}\left(1 + \frac{\alpha^2 m_e^2}{4m_\chi^2}\right)\!\left(1 + \frac{\alpha^2}{4}\right), \qquad \mu_{\chi e} = \frac{m_\chi m_e}{m_\chi + m_e}.
    \label{eq:sigmabar_e}
\end{equation}

\begin{table}[htbp]
    \centering
    \caption{Experimental inputs and $90\%$ C.L. upper limits used in the statistical recasting procedure~\eqref{eq:N_sig}--\eqref{eq:g_solve}, together with the resulting exact and static-numerator coupling bounds $\sqrt{g_\chi g_e}^{\mathrm{max}}$ at $m_\chi = 10\,\mathrm{keV}$ and $m_S = 10\,\mathrm{MeV}$ ($D_{\mathrm{eff}} = 1\,\mathrm{kpc}$).}
    \label{tab:exp_inputs}
    \resizebox{\textwidth}{!}{%
    \begin{tabular}{lcccccc}
        \toprule
        Experiment & Exposure $M_{\mathrm{det}}\Delta t$ & Electron Exposure $N_e \Delta t\,[\mathrm{e\cdot s}]$ & Window $[T_e^{\mathrm{th}}, T_e^{\mathrm{upper}}]$ & Efficiency $\bar{\epsilon}_{\mathrm{det}}$ & $N_{\mathrm{sig}}^{90\%}$ & $\sqrt{g_\chi g_e}^{\mathrm{max}}$ ($10\,\mathrm{keV}$, Exact / Static) \\
        \midrule
        Super-Kamiokande IV\cite{SuperK_BDM2023} & $161.9\,\mathrm{kton\cdot yr}$ & $1.70 \times 10^{42}$ & $0.1\text{--}10\,\mathrm{GeV}$ & $0.75$ & $28.5$ & $2.39 \times 10^{-2}$ / $9.91 \times 10^{-1}$ \\
        XENONnT\cite{XENONnT2023}                & $1.16\,\mathrm{ton\cdot yr}$   & $6.72 \times 10^{36}$ & $1\text{--}30\,\mathrm{keV}$   & $0.85$ & $12.4$ & $8.16 \times 10^{-1}$ / $1.92 \times 10^{0}$ \\
        PandaX-4T\cite{PandaX4T_BDM2022}         & $0.63\,\mathrm{ton\cdot yr}$   & $3.65 \times 10^{36}$ & $1\text{--}30\,\mathrm{keV}$   & $0.82$ & $10.8$ & $8.79 \times 10^{-1}$ / $2.06 \times 10^{0}$ \\
        \bottomrule
    \end{tabular}%
    }
\end{table}

\begin{figure}[htbp]
    \centering
    \includegraphics[width=0.95\textwidth]{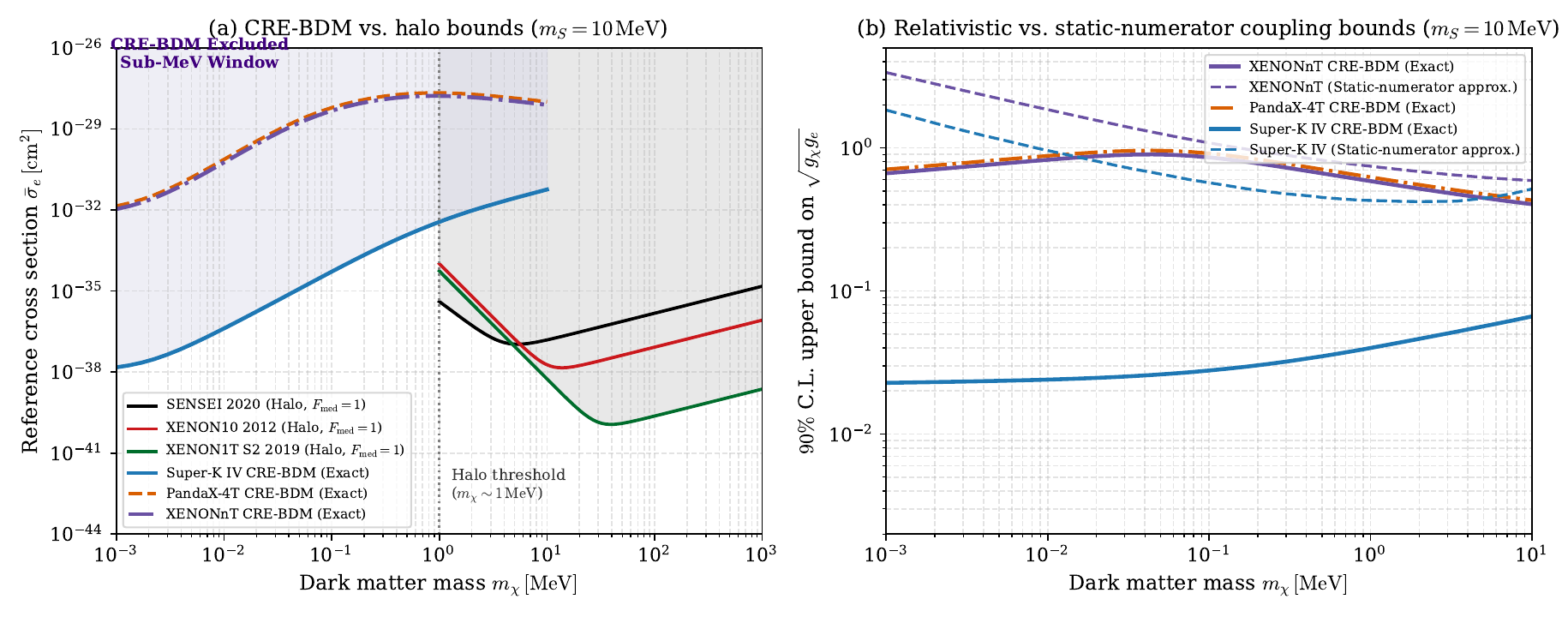}
    \caption{(a) $90\%$ C.L. exclusion bounds in the $(m_\chi, \bar{\sigma}_e)$ plane across $1\,\mathrm{keV} \le m_\chi \le 1\,\mathrm{GeV}$ for $m_S = 10\,\mathrm{MeV}$ and $D_{\mathrm{eff}} = 1\,\mathrm{kpc}$, comparing CRE-BDM constraints in the $\mathrm{keV}$--$10\,\mathrm{MeV}$ range from Super-Kamiokande IV~\cite{SuperK_BDM2023}, XENONnT~\cite{XENONnT2023}, and PandaX-4T\cite{PandaX4T_BDM2022} against published halo bounds ($F_{\mathrm{med}} = 1$) from SENSEI 2020~\cite{SENSEI2020}, XENON10 2012~\cite{XENON10_2012}, and XENON1T S2 2019~\cite{XENON1T_S2_2019}. (b) Comparison of the $90\%$ C.L. upper bounds on $\sqrt{g_\chi g_e}$ in the $\mathrm{keV}$--$10\,\mathrm{MeV}$ mass range ($m_S = 10\,\mathrm{MeV}$) obtained from the exact relativistic spinor amplitude (solid and dash-dotted curves) versus the static-numerator approximation (dashed curves).}
    \label{fig:mpla_fig3}
\end{figure}

Figure~\ref{fig:mpla_fig3}(a) presents the $90\%$ C.L. exclusion contours in the $(m_\chi, \bar{\sigma}_e)$ plane for the benchmark mediator mass $m_S = 10\,\mathrm{MeV}$. We emphasize an important condition for comparing CRE-BDM bounds and published halo bounds on the same $\bar{\sigma}_e$ plane: at $m_S = 10\,\mathrm{MeV} \gg \alpha m_e \simeq 3.73\,\mathrm{keV}$, the halo mediator form factor satisfies $F_{\mathrm{med}}(q) = (\alpha^2 m_e^2 + m_S^2)/(q^2 + m_S^2) = 1$ to within $0.1\%$ for all halo momentum transfers, so the published $F_{\mathrm{med}} = 1$ halo exclusion curves from SENSEI, XENON10, and XENON1T apply directly without recasting; for lighter mediators ($m_S \lesssim \alpha m_e$), halo limits cannot be converted by Eq.~\eqref{eq:sigmabar_e} alone without recasting the detector ionization form factor~\cite{Xuan2026_CIP}. While halo searches constrain $m_\chi \gtrsim 1\,\mathrm{MeV}$, CRE-BDM searches cover the $\mathrm{keV}$--$10\,\mathrm{MeV}$ range and extend sensitivity into the sub-MeV window ($1\,\mathrm{keV} \le m_\chi < 1\,\mathrm{MeV}$), with Super-Kamiokande IV providing the strongest bound ($\sqrt{g_\chi g_e} \le 2.28 \times 10^{-2}$ at $m_\chi = 1\,\mathrm{keV}$) owing to its $161.9\,\mathrm{kton\cdot yr}$ exposure and the strong $q^4$ spinor-recoil enhancement at $T_e \ge 100\,\mathrm{MeV}$.

Figure~\ref{fig:mpla_fig3}(b) and Table~\ref{tab:exp_inputs} compare the coupling bounds $\sqrt{g_\chi g_e}^{\mathrm{max}}$ obtained using the exact relativistic cross sections against those from the static-numerator approximation. In XENONnT ($T_e \in [1, 30]\,\mathrm{keV}$), where relativistic enhancement occurs primarily during cosmic-ray upscattering, the static-numerator approximation under-constrains $\sqrt{g_\chi g_e}$ by a factor of $5.06$ at $m_\chi = 1\,\mathrm{keV}$ and $2.35$ at $m_\chi = 10\,\mathrm{keV}$. In Super-Kamiokande IV ($T_e \in [0.1, 10]\,\mathrm{GeV}$), where both upscattering and detection occur at $|t| \gg 4m_e^2, 4m_\chi^2$, omitting the relativistic spinor recoil terms weakens the inferred coupling bound by a factor of $41.5$ at $m_\chi = 10\,\mathrm{keV}$ and $80.6$ at $m_\chi = 1\,\mathrm{keV}$.

\section{Conclusions}\label{sec:conclusions}
In this work, we have studied relativistic elastic scattering of Dirac dark matter with charged leptons through a real scalar mediator $S$ in the low-energy mass-eigenbasis framework and applied it to cosmic-ray electron boosted dark matter (CRE-BDM). Our main results are:
\begin{enumerate}\setlength{\itemsep}{1pt}\setlength{\parskip}{0pt}
    \item We formulated the exact two-stage laboratory-frame relativistic kinematics for galactic cosmic-ray electron upscattering and terrestrial detector scattering, identifying the distinct momentum-transfer thresholds for crossing the electron ($4m_e^2$) and dark matter ($4m_\chi^2$) mass scales and proving the analytical condition~\eqref{eq:df_dq2} under which the recoil spectrum rises with $T_e$.
    \item We decomposed the tree-level cross section into the static-numerator contribution ($\sigma_0$) and the linear ($\sigma_1$) and quadratic ($\sigma_2$) spinor-recoil contributions, verified the exact power-series representation~\eqref{eq:Phi_series} against 80-digit reference calculations to $\le 1.2 \times 10^{-15}$ relative accuracy across multiple mass and energy regimes (Table~\ref{tab:validation}), and showed that $\sigma_1 + \sigma_2$ begins to exceed $\sigma_0$ at $T_\chi \sim m_\chi$ ($\sigma_{\chi e}/\sigma_0 = 2.241$ at $T_\chi = 10\,\mathrm{MeV}$) and enhances the cross section by factors of $27.1$ to $2.47 \times 10^4$ for $T_\chi \in [0.1, 100]\,\mathrm{GeV}$.
    \item Using the CM-frame helicity amplitudes and $J = 0$ Jacob--Wick projection, we showed that the dynamical scalar propagator bounds $|a^0_{00}(s)| \to g_\chi g_\ell / (16\pi)$ and yields a high-energy cross section decreasing as $g_\chi^2 g_\ell^2/(16\pi s)$, avoiding the linear $\mathcal{O}(s/m_S^4)$ unitarity violation of the EFT contact approximation.
    \item Within the effective-propagation CRE-BDM model ($D_{\mathrm{eff}} = 1\,\mathrm{kpc}$), we documented the experimental inputs and statistical procedure (Table~\ref{tab:exp_inputs}) to derive $90\%$ C.L. exclusion bounds in the $\mathrm{keV}$--$10\,\mathrm{MeV}$ mass range from Super-Kamiokande IV, XENONnT, and PandaX-4T, demonstrating that retaining the full relativistic spinor recoil structure strengthens the sub-MeV coupling bounds by factors of $2.4\text{--}5.1$ in XENONnT/PandaX-4T and $20\text{--}80$ in Super-Kamiokande compared to the static-numerator approximation.
\end{enumerate}

In subsequent works, we will extend this framework to thermal freeze-out relic abundance ($\Omega_\chi h^2$) via $s$-channel annihilation $\chi\bar{\chi} \to S \to \ell^+\ell^-$ and one-loop renormalized quantum corrections ($S\gamma\gamma$, $SZZ$, and loop-induced nucleon scattering).

\section*{Acknowledgements}
This work was completed building upon scientific materials left by the late Professor Nguyen Suan Han. The author gratefully acknowledges his mentorship, guidance, and lasting scientific influence. Sincere thanks are also extended to the Faculty of Physics and Chemical Engineering, Le Quy Don Technical University for academic support.

\end{document}